\documentclass[journal]{IEEEtran}
\usepackage{multirow}

\usepackage[sorting=none,maxbibnames=99]{biblatex}
\usepackage{color, colortbl}
\usepackage{amsfonts} 
\usepackage{amssymb}
\usepackage{breqn}

\usepackage{algorithmic}
\usepackage[dvipsnames]{xcolor}
\usepackage{amsmath}
\usepackage{array}
\usepackage{makecell}
\usepackage{mathtools}

\makeatletter
\newcommand{\thickhline}{%
    \noalign {\ifnum 0=`}\fi \hrule height 1pt
    \futurelet \reserved@a \@xhline}

\usepackage{algorithm}
\usepackage{algorithmic}
\makeatletter
\newcommand\fs@norules{\def\@fs@cfont{\bfseries}\let\@fs@capt\floatc@ruled
  \def\@fs@pre{}%
  \def\@fs@post{}%
  \def\@fs@mid{\kern3pt}%
  \let\@fs@iftopcapt\iftrue}
\makeatother
\floatstyle{norules}
\restylefloat{algorithm}

\begin{document}
%

\title{Seizing Opportunity: Maintenance Optimization in Offshore Wind Farms Considering Accessibility, Production, and Crew Dispatch }

\author{Petros~Papadopoulos, David~W.~Coit and
         Ahmed~Aziz~Ezzat$^\dagger$
\thanks{P. Papadopoulos, D. Coit and A. Ezzat are with the Department of Industrial \& Systems Engineering, Rutgers University, NJ, 08854, USA.}
\thanks{$^\dagger$Corresponding author; 
contact e-mail: aziz.ezzat@rutgers.edu}
}
%
%

\markboth{This work has been accepted for publication at the IEEE Transactions on Sustainable Energy}
{Papadopoulos \MakeLowercase{\textit{et al.}}: Bare Demo of IEEEtran.cls for IEEE Journals}
%



\maketitle

\begin{abstract}
Operations and Maintenance (O\&M) constitute a major contributor to offshore wind's cost of energy. Due to the harsh and remote environment in which offshore turbines operate, 
there has been a growing interest in \textit{opportunistic} maintenance scheduling for offshore wind farms, wherein grouping maintenance tasks is incentivized at times of opportunity. Our survey of the literature, however, reveals that there is no unified consensus on what constitutes an ``opportunity'' for offshore maintenance. We therefore propose an opportunistic maintenance scheduling approach which defines an opportunity as either crew-dispatch-based (initiated by a maintenance crew already dispatched to a neighboring turbine), production-based (initiated by projected low production levels), or access-based (initiated by a provisionally open window of turbine access). We formulate the problem as a multi-staged rolling-horizon mixed integer linear program, and propose an iterative solution algorithm to identify the optimal hourly maintenance schedule, which is found to be drastically different, yet substantially better, than those obtained using offshore-agnostic strategies. Extensive numerical experiments on actual wind, wave, and power data demonstrate substantial margins of improvement achieved by our proposed approach, across a wide variety of key O\&M metrics.
\end{abstract}


\IEEEpeerreviewmaketitle

\section{Introduction}
\IEEEPARstart{T}{he} global capacity of offshore wind energy is projected to increase by at least $15$ times by $2040$, ranging between $350$ GW and $560$ GW \cite{iea2019outlook_OW}. Despite the promising outlook, the offshore wind industry still faces substantial challenges pertaining to the high expenditures of operating and maintaining offshore wind farms. Unlike their onshore counterparts, offshore turbines operate at relatively unexplored territories, higher altitudes, less accessible locations, and under harsher weather and marine conditions, ultimately inflating the cost of offshore maintenance operations \cite{ren2021offshore}. 
With O\&M currently contributing about $30$\% of offshore wind's cost of energy \cite{stehly20202018}, a challenge of central interest to the offshore wind industry is \textit{how to establish a maintenance scheduling approach which is tailored to offshore wind farm operations and their offshore-specific operational and environmental conditions?}

While there exists a large body of literature on wind farm maintenance scheduling optimization 
\cite{byon2010season,lei2018maintenance}, offshore maintenance in particular entails a set of unique offshore-specific challenges, including: (i) \emph{High crew dispatch costs}: Offshore turbines are often installed at remote locations where wind conditions are most favorable, making the cost of assembling and transporting a maintenance crew considerably high; (ii) \emph{High production losses}: With the increasing scale and capacity of offshore turbines, the cost of downtime is becoming too substantial to tolerate due to the associated production losses of a failing ultra-scale offshore turbine; and (iii) \emph{Limited accessibility}: Harsh weather and marine conditions can frequently prohibit turbine access for long periods of time, which can range from a few hours up to several days.

Due to the aforementioned offshore-specific challenges, there has been a growing interest in \textit{opportunistic} maintenance scheduling, wherein grouping of maintenance tasks is incentivized at times of opportunity. We note, however, that the exact definition of what constitutes a maintenance opportunity is subjective among researchers \cite{ren2021offshore}. Perhaps the most prevalent definition of a maintenance opportunity is the idea of grouping maintenance tasks for turbines that are of spatial proximity to each other, thereby avoiding additional costs incurred by assembling and dispatching a crew at two separate maintenance occasions. With state-of-the-art estimates suggesting that crew dispatch/transport costs contribute about $50$-$73\%$ of offshore maintenance expenses \cite{dalgic2015offshorelogistics, carroll2016failure}, efforts devoted towards offsetting these costs are expected to substantially drive down the total O\&M expenditures. We denote the set of approaches which adopt this view of opportunistic maintenance as \textit{dispatch-based} opportunistic strategies \cite{shafiee2015opportunistic, wang2019optimizing, ding2012opportunistic, lu2017opportunistic, sarker2016minimizing, yildirim2017integrated}. \par 

A less formally adopted, albeit rational approach for opportunistic maintenance is to leverage the opportunities initiated by anticipated low production profiles in an attempt to minimize the production losses due to maintenance shutdowns. In other words, if scheduling a maintenance is imminent, then an operator may prefer to schedule the maintenance at times when power production is projected to be low, thus avoiding a ``lost opportunity cost'' represented in the forfeited market economic value. This is extremely relevant in light of the emerging ultra-scale turbines ($\geq$ $12$ MW) which will soon be deployed at different regions across the world \cite{GE}. Different from dispatch-based strategies, we denote this set of efforts as \textit{production-based} opportunistic strategies \cite{besnard2009optimization, besnard2011stochastic,4463801, yang2020operations}. 

Lastly, few studies consider opportunities created by provisionally open windows of turbine access initiated by tolerable weather/marine circumstances. Our analysis of actual offshore data suggests that an offshore turbine can be inaccessible for $53$\% of its operational time, with up to $6$ days of consecutive constrained access. Understandably, inaccurate or overlooked access information in maintenance scheduling can lead to aborted/delayed maintenance activities, cascading failures, and prolonged downtimes. In contrast to the first two groups which define opportunistic maintenance as dispatch- or production-based, this set of methods is referred to hereinafter as \textit{accessibility-based} opportunistic strategies \cite{taylor2018probabilistic, lubing2019opportunistic, zhang2021uncertain}.  \par 

In light of the above-referenced works, we summarize our contributions as follows: we propose a holistic approach for opportunistic maintenance scheduling in offshore wind farms\textemdash one which embodies all the aforementioned offshore-specific considerations.
In pursuit of that goal, we propose a mathematical formulation based on a mixed integer linear program (MILP) with a two-staged rolling time horizon, which is inspired by the general maintenance procedure followed in the offshore wind industry. We further propose an iterative solution algorithm which yields the optimal day-ahead hourly maintenance schedule, by harnessing the joint opportunities created by crew dispatch, production, and accessibility. 

A major takeaway of this article is to shed light on the need to develop offshore-tailored O\&M solutions which fundamentally depart from their onshore counterparts. Our numerical experiments, using actual wind, wave, and power data, provide valuable insights into the economic impacts of each of the three opportunistic considerations. Those experiments further assert that \textit{holistically} accounting for those offshore-specific considerations \textit{in tandem}, yields drastically different (yet better) schedules relative to those obtained by strategies that may be well-suited for onshore wind farm maintenance. As such, we hereinafter denote our approach as the \textit{\underline{h}olistic \underline{o}pportunity-based \underline{st}rategy} (HOST). The HOST method is tractable (ensuring easy adoption by practitioners), effective (as shown by large O\&M cost reductions), scalable (solved within seconds), and versatile (easily extensible to accommodate different O\&M conditions/parameters). 

Before we conclude this section, we would like to draw a complete picture of where our proposed approach fits within the current industrial practice. It is important to note, however, that the industrial practice in offshore maintenance scheduling can largely vary across different operators and O\&M service providers, and is often based on in-house experience. Nevertheless, a broadly followed procedure is to adopt a multi-staged structure with a longer-term planning stage wherein work orders are identified and prioritized, which in turn informs a shorter-term operational stage, wherein the day-ahead maintenance schedule is decided at the end of the working day, following a debrief meeting between the technicians and operations control team, and in light of updated weather forecasts and O\&M information. The day-ahead maintenance schedule includes a detailed timetable of the high-priority maintenance tasks \cite{koltsidopoulos2020data, kolios2018deliverable}. The HOST method proposed herein faithfully follows the above-mentioned structure, by providing an hourly schedule for the day-ahead operational stage, together with a daily schedule for longer-term tasks. A recent report by New York State Energy Research and Development Authority (NYSERDA) surveying the state-of-the-art in the offshore wind industry explicitly states that O\&M costs are significantly more expensive in offshore environments mainly due to accessibility and logistics \cite{nyserda2019}. This, in fact, is the driving motivation of HOST, with its ultimate objective of reducing offshore O\&M expenditures.

The remainder of the paper is structured as follows. In Section II, we begin by outlining the main assumptions of our problem setting, then proceed to present the mathematical formulation and solution procedure. Section III describes the datasets used in our study, the experimental setups, and maintenance benchmarks. Qualitative and quantitative analysis on two case studies is then conducted in Section IV to highlight the effectiveness of our proposed approach. Finally, Section V concludes the paper and highlights future research directions. 

\section{Mathematical Model}
We begin this section by listing the key assumptions of HOST, followed by its mathematical formulation. All variables, parameters and sets have been declared in Table \ref{tab:nom}. We then propose an iterative solution algorithm to efficiently find a cost-minimal, offshore-tailored 
maintenance schedule.

\subsection{Model assumptions}
    ($\mathcal{A}1$) An offshore wind farm consists of a set of $I$ wind turbines (WTs), each requiring a maintenance of $\tau_i$ hours, which has to be performed within a period of $J$ days, where $i \in \mathcal{I} = \{1, ..., I\}$ is the turbine index. The type of maintenance tasks considered in this work are primarily minor to medium repairs which typically require less than a day to complete \cite{dinwoodie2015reference}. Examples of those tasks include manual resets, minor repairs to the pitch and hydraulic systems (e.g., oil leaks, pump issues), electric, and electronic components. Those tasks constitute the majority of maintenance actions in a wind farm ($\sim$\hspace{-.1mm}$75$\% as reported in \cite{carroll2016failure}). While not considered in this paper, our model is potentially extensible to accommodate multi-type maintenance tasks (e.g., minor to major repairs) with careful adaptations. (e.g., extending the planning horizon $J$, redefining access windows and maintenance crew/vessel parameters, adjusting repair times, and failure rates). 
    
    ($\mathcal{A}2$) Each turbine has a residual life estimate (RLE) of $\lambda_i$ days, known to the operator \textit{a priori}. In practice, $\lambda_i$ can be informed by condition monitoring systems, often combined with expert judgement. A maintenance action is assumed to be preventive (PM) if performed before the RLE, and corrective (CM) otherwise. Furthermore, in addition to ``scheduled'' tasks, unexpected failures may occur to a subset of WTs, triggering additional CM actions. Understandably, we associate a higher maintenance cost with CM actions, since a failed component often requires a full repair/replacement, notwithstanding potential cascading failures for other components in its proximity. 
    
    ($\mathcal{A}3$) The optimization horizon is divided into two sets: the short-term horizon (STH) and the long-term horizon (LTH). The STH has an hourly resolution and spans a duration of one day (i.e., day-ahead planning). The LTH has a daily resolution, starting from the day after the STH, up to $J$ days ahead. The indices $t$ and $d$ denote the $t$-th hour and $d$-th day, while the sets $\mathcal{T}$ and $\mathcal{D}$ denote the sets of STH hours and LTH days, respectively. A superscript $L$ denotes a variable/parameter associated with the LTH, while the same variable/parameter, without the superscript, pertains to the STH.
    
    ($\mathcal{A}4$) For both the STH and LTH, point forecasts of wind speed ($\mbox{V}_{t,i}, \mbox{V}_{d,i}^L$) and wave height ($\mbox{H}_{t,i}, \mbox{H}_{d,i}^L$) are available to the operator \textit{a priori}. In practice, those forecasts can be obtained via meso-scale numerical weather prediction (NWP) models \cite{kumler2019comparison} (especially for longer horizons), statistical/machine learning models \cite{ezzat2018spatio, ezzat2020turbine} for shorter time horizons, or a combination of both. Those forecasts will be primarily used to predict the power profiles and assess turbine access\textemdash more details in Section III. 
    
     ($\mathcal{A}5$) Turbines are accessed by crew transport vessels (CTVs) which are subject to accessibility constraints defined by wind speed and wave height safety thresholds, denoted by $\nu$ (m/s) and $\eta$ (m), respectively. A turbine is accessible in the STH when the weather/marine conditions are within the safety limits for $\tau_i$ consecutive hourly intervals to ensure an uninterrupted completion of the maintenance task. For the LTH, the turbine is deemed accessible if there is at least one opportunity to schedule a maintenance action during the $d$-th day. Furthermore, turbines are only accessible during daylight, via daily vessel rentals from a private O\&M contractor.
    
\subsection{Mathematical formulation}

\subsubsection*{Objective function}
The objective, expressed in (\ref{eq1}), is to maximize the total profit over the optimization horizon, which is the sum of the short-term profit $s$ and the long-term profits $\{l_d\}_{d\in\mathcal{D}}$. The set of decision variables constitute a collection of PM and CM binary variables indicating whether a particular maintenance type is scheduled for a turbine site at a particular time. For the STH, these are denoted by $\{m_{t,i}, n_{t,i}\}$ $\forall i, t$ and for the LTH, they are denoted by $\{m_{d,i}^L, n_{d,i}^L\}$ $\forall i, d$. 
\begin{equation}\label{eq1}
    \max \hspace{0.25cm} \bigg\{s + \sum_{d\in \mathcal{D}} l_d\bigg\}
\end{equation}
The short-term profit, $s$, defined in (\ref{eq:eq2}), is the difference between the revenue from the power produced at the $t$-th hour and $i$-th turbine, $p_{t,i}$ (MWh)\textemdash sold at the hourly market price $\Pi_t~(\$/\mbox{MWh})$\textemdash and the maintenance costs, which are dictated by the PM and CM cost coefficients, K $(\$/\mbox{task})$ and $\Phi~(\$/\mbox{task})$, respectively. Crew-related costs are calculated based on the crew hourly cost rate $\Psi~(\$/\mbox{h})$, the daily vessel rental cost $\Omega~(\$/\mbox{day})$, and the crew overtime cost rate $\mbox{Q}~(\$/\mbox{h})$. The integer variable $q \in \mathbb{Z}^+$ denotes the overtime hours worked in the STH, while the binary variables $x_{t,i}$ and $v$ denote whether a turbine is under maintenance, and whether a vessel is rented during the STH, respectively. Deriving $p_{t,i}$, $x_{t,i}$, $v$, $q$ using the decision variables will follow in the sequel.
\begin{dmath}\label{eq:eq2}
    s=\hspace{-0.45cm}\sum_{i \in \mathcal{I} , t \in \mathcal{T}}\hspace{-0.3cm}[\Pi_t \cdot p_{t,i} - \mbox{K} \cdot m_{t,i} - \Phi \cdot n_{t,i} - \Psi \cdot x_{t,i}] {- \Omega \cdot v - \mbox{Q} \cdot q}
\end{dmath}

The long-term daily profit, in (\ref{eq:eq3}), is similarly calculated, with the exception that there is no overtime since hourly tracking is not possible in the LTH. The daily electricity selling price $\Pi_d$ is calculated by averaging hourly prices at day $d$. The crew work hours are calculated as the product of the maintenance time $\tau_i$ by the number of maintenance tasks scheduled at that day, $m_{d,i}^L+n_{d,i}^L$. The variable $v_d^L \in \{0,1\}$ denotes whether a vessel is rented at the $d$-th day of the LTH. 
\begin{dmath} \label{eq:eq3}
 l_d = \sum_{i \in \mathcal{I}} \bigg[\Pi_d \cdot p_{d,i}^L - \mbox{K} \cdot m_{d,i}^L - \Phi \cdot n_{d,i}^L - {\Psi\cdot\tau_i (m_{d,i}^L+n_{d,i}^L)\bigg] - \Omega \cdot v_d^L \quad \forall d \in \mathcal{D}}
\end{dmath}

\subsubsection*{Maintenance constraints} 
The inequality in (\ref{eq:maint}) forces a maintenance task, either PM or CM, to be scheduled either in the STH or the LTH as long as the maintenance requirement parameter, denoted by $\theta_i \in \{0,1\}$, is set to $1$. This parameter is an input to the optimization, denoting whether maintenance is required for the $i$-th turbine. 
\begin{equation} \label{eq:maint}
\sum_{t \in \mathcal{T}}(m_{t,i}+n_{t,i})+\sum_{d \in \mathcal{D}}(m_{d,i}^L+n_{d,i}^L) \geq \theta_i \quad \forall  i \in \mathcal{I}
\end{equation}

The maintenance crew, once dispatched, will be occupied for $\tau_i$ consecutive hourly intervals. This is ensured by the constraint in (\ref{eq:crew}). An upper bound on the number of available maintenance crews is set to $\mbox{B}$ (crews), as shown in (\ref{eq:crew2}). 
\begin{equation} \label{eq:crew}
\sum_{\tilde{t}=t}^{t+\tau_i -1} x_{\tilde{t},i} \geq \tau_i \cdot (m_{t,i}+n_{t,i}) \quad \forall t \in \mathcal{T}, i \in \mathcal{I}
\end{equation}
\begin{equation} \label{eq:crew2}
\sum_{i \in \mathcal{I}} x_{t,i} \leq \mbox{B} \quad \forall t \in \mathcal{T}
\end{equation}

If not preventively maintained on or before their RLEs, turbines fail and will require a corrective maintenance. This is expressed in (\ref{eq:dead1}) and (\ref{eq:dead3}) for STH, and (\ref{eq:dead2}) and (\ref{eq:dead4}) for LTH. 

\begin{equation} \label{eq:dead1}
m_{t,i} \leq \frac{24 \cdot \lambda_i}{t} \quad \forall t \in \mathcal{T}, i \in \mathcal{I}
\end{equation}
\begin{equation}\label{eq:dead2}
    m_{d,i}^L \leq \frac{ \lambda_i}{d} \quad \forall d \in \mathcal{D}, i \in \mathcal{I}
\end{equation}
\begin{equation}\label{eq:dead3}
    n_{t,i} \leq \frac{t}{24 \cdot \lambda_i + \beta} \quad \forall t \in \mathcal{T}, i \in \mathcal{I}
\end{equation}
\begin{equation}\label{eq:dead4}
n_{d,i}^L \leq \frac{d_J - \lambda_i}{d_J -d+ \beta}  \quad \forall d \in \mathcal{D}, i \in \mathcal{I},
\end{equation}
where $d_J \in \mathcal{D}$ is the last element of the set $\mathcal{D}$, while $\beta$ is a small value to ensure numerical stability when $d=d_J$ and to enforce a value of $1$ for $n_{t,i}$ and $n_{d,i}^L$ only when $t > 24 \cdot \lambda_i$ and $d > \lambda_i$, respectively.

\subsubsection*{Turbine availability}
A failed turbine (i.e., one which has not been maintained at or before its RLE) remains unavailable until a CM action is performed. In case of the STH, this can be expressed by the following inequality:
\begin{dmath} \label{eq:availability}
y_{t,i} \leq \left( \frac{24 \cdot \lambda_i}{t} \right)^{\mbox{M}}+ \frac{24 \cdot \sum_{t \in \mathcal{T}}n_{t,i}- \sum_{t \in \mathcal{T}}(t \cdot n_{t,i})}{t_F-t+\beta}+{1-\theta_i,
\quad \forall t \in \mathcal{T}, i \in \mathcal{I}}, \end{dmath}
where $t_F$ is the last element of the set $\mathcal{T}$. The first term of the right-hand side (RHS) is raised to an arbitrary big value, M, so that it rapidly drops to $0$ if $t > 24 \cdot \lambda_i$, i.e. when the time index of the STH exceeds the RLE, $\lambda_i$, which is given in days. If the turbine does not require maintenance, then we have $\theta_i=0$, and the last term of the RHS, namely $1-\theta_i$ will be 1, signifying an operational turbine. In case the RLE is reached, and the turbine still requires maintenance ($\theta_i=1$), then the turbine fails, and can only return to its former operational status once a CM action is performed. This is enforced by the middle term in the RHS; if no CM is scheduled, then $n_{t,i}=0$ and the term drops to 0. In contrast, if $n_{t,i}=1$, the term is greater than $1$ after the time of maintenance. A similar constraint for the LTH is shown in~(\ref{eq:availability2}). 
\begin{dmath} \label{eq:availability2}
y_{d,i}^L \leq \left( \frac{ \lambda_i}{d} \right)^{\mbox{M}}\hspace{-0.05cm}+\frac{d_J- \sum_{d \in \mathcal{D}}(d \cdot n_{d,i}^L)}{d_J-d+\beta} \hspace{0.3cm} {\forall d \in \mathcal{D}, i \in \mathcal{I}} \end{dmath}
As expressed in (\ref{eq:maintt}), a turbine under maintenance remains unavailable until the task is completed.
\begin{equation} \label{eq:maintt}
   y_{t,i} \leq 1 - x_{t,i} \quad \forall t \in \mathcal{T}, i \in \mathcal{I}
\end{equation}

\subsubsection*{Vessel rental}
Vessels are rented per day if a maintenance task is scheduled during that day. This is enforced via (\ref{eq:vessel1})-(\ref{eq:vessel2}), for the STH and LTH, respectively.
\begin{equation} \label{eq:vessel1}
    v \geq m_{t,i}+n_{t,i} \quad \forall t \in \mathcal{T}, i \in \mathcal{I}
\end{equation}
\begin{equation}\label{eq:vessel2}
  v_d^L \geq m_{d,i}^L+n_{d,i}^L \quad \forall d \in \mathcal{D}, i \in \mathcal{I}  
\end{equation}

\subsubsection*{Overtime}
The total work hours, as shown in (\ref{eq:workhours}), cannot exceed $\mbox{W}$ (h/crew). Otherwise, overtime hours, tracked by $q$, are incurred and compensated at a higher rate determined by $\mbox{Q}$ (\$/h). For the LTH, the total number of hours worked in a day by each crew is limited to $\mbox{W}$ hours, as shown in (\ref{eq:workhours2}). As mentioned earlier, no overtime is allowed in the LTH. 
\begin{equation} \label{eq:workhours}
\sum_{t \in \mathcal{T}, i \in \mathcal{I}} x_{t,i} \leq \mbox{B} \cdot \mbox{W}+q
\end{equation}
\begin{equation} \label{eq:workhours2}
\sum_{i \in \mathcal{I}} (m_{d,i}^L+n_{d,i}^L) \cdot \tau_i \leq \mbox{B} \cdot \mbox{W} \quad \forall d \in \mathcal{D}
\end{equation}

\subsubsection*{Power level}
The hourly power output, $p_{t,i}$, is computed as a fraction $f_{t,i} \in [0,1]$ of the turbine's rated capacity R (MW), multiplied by the turbine's availability $y_{t,i}$, as shown in (\ref{eq:eqnom}). When the turbine is in a failed state, or under maintenance, the power output of the wind turbine is zero as the availability variable, $y_{t,i}$, is also zero, therefore, the operator forfeits the associated revenue. The fraction $f_{t,i}$ is called \textit{the normalized power level} and is a function of the hub-height hourly wind speed at the turbine's location, $\mbox{V}_{t,i}$ (m/s). The exact procedure to calculate $f_{t,i}$ given wind forecasts is deferred to Section III. 
\begin{equation}\label{eq:eqnom}
    p_{t,i}=\mbox{R} \cdot f_{t,i} \cdot y_{t,i} \quad \forall t \in \mathcal{T}, i \in \mathcal{I}
\end{equation}
Likewise, the daily power output, $p_{d,i}^L$, is defined in (\ref{eq:nom2}), wherein $f_{d,i}^L \in [0, 1]$ is a function of the daily average wind speed at the turbine's location, $\mbox{V}_{d,i}^L$ (m/s). The term $m_{d,i}^L \cdot \tau_i /24$, accounts for the production loss, in case a PM is scheduled at the $d$-th day for the $i$-th turbine in the LTH.
\begin{equation} \label{eq:nom2}
p_{d,i}^L=24 \cdot \mbox{R} \cdot f_{d,i}^L \cdot \left(y_{d,i}^L- m_{d,i}^L \cdot \frac{\tau_i}{24} \right) \quad \forall d \in \mathcal{D}, i \in \mathcal{I}
\end{equation}
\subsubsection*{Curtailment} Curtailing production is accounted for by introducing the parameter $\mbox{C}_t\in [0,1]$ in (\ref{eq:curtailment}) which defines the fraction of the farm-level power output that can be sold to the grid at each hourly interval $t$. 
\begin{equation} \label{eq:curtailment}
\sum_{i \in \mathcal{I}} p_{t,i} \leq \sum_{i \in \mathcal{I}} \cdot f_{t,i} \cdot \mbox{R} \cdot \mbox{C}_t \quad \forall t \in \mathcal{T}
\end{equation}
\subsubsection*{Accessibility}
The parameters $\alpha_{t,i} \in \{0,1\}$ and $\alpha_{d,i}^L \in \{0,1\}$ define the accessibility at a specific location and time. Access calculations, given wind speed and wave height forecasts, are made offline prior to the optimization (More details in Section III). In (\ref{eq:access1}) and (\ref{eq:access2}), a PM or a CM action is only scheduled in the STH if $\alpha_{t,i} = 1$ and in the LTH if $\alpha_{d,i}^L = 1$.
\begin{equation} \label{eq:access1}
    m_{t,i}+n_{t,i} \leq \alpha_{t,i} \quad \forall t \in \mathcal{T}, i \in \mathcal{I}
\end{equation}
\begin{equation} \label{eq:access2}
    m_{d,i}^L+n_{d,i}^L \leq \alpha_{d,i}^L \quad \forall d \in \mathcal{D}, i \in \mathcal{I}
\end{equation}

Next, given the model formulation, we describe a rolling horizon solution procedure to efficiently find and update the optimal hourly maintenance schedule. 

\subsection{The rolling horizon adaptation and solution algorithm}
Solving the model in Section II-B yields an hourly schedule for the $1$-day ahead STH and a daily schedule for the $J$-days ahead LTH. In practice, we are interested in continuously updating the schedule as time progresses and more information is revealed. Towards that, we propose a rolling-horizon adaptation of our model, along with an iterative solution procedure, which is presented in Algorithm 1. Each iteration of this procedure entails solving one instance of the MILP model, wherein a subset of the model indices and parameters are updated at the start of each iteration based on the solution of the previous iteration and the newly revealed information. \par 

Specifically, before each iteration, we shift the starting times of the STH and LTH by $24$ hours and $1$ day, respectively. Weather-dependent parameters (e.g. accessibility, power production) are updated at the start of each run in light of the newly revealed forecast information. Two consecutive iterations are coupled by updating the values of $\{\theta_i\}_{i \in \mathcal{I}}$: if the $i$-th turbine has been maintained in the STH of the previous iteration, we set $\theta_i = 0$, indicating that the turbine no longer requires maintenance. At the end, the individual STH solutions from all iterations are ``concatenated'' to produce the final hourly schedule for the whole optimization horizon. This constitutes our final hourly schedule. 

 \begin{algorithm}[]
 \caption{Iterative solution procedure for HOST}
  \renewcommand{\arraystretch}{1.4}
  \begin{algorithmic}[1]
  \STATE \textit{Input} turbine maintenance parameters $\{\lambda_i, \theta_i, \tau_i\}_{i \in \mathcal{I}}$
   \STATE \textit{Input} cost and price parameters $\mbox{K}, \Phi, \Psi, \Omega, \mbox{Q}, \Pi_t, \Pi_d, \mbox{C}_t$
   \STATE \textit{Input} operational parameters $\mbox{R}, \nu, \eta, \mbox{B}, \mbox{W}$
   \STATE \textit{Input} the optimization horizon $J$, where $J > \underset{i}{\max\hspace{1mm}}{\lambda_i}$  \\
   \STATE \textit{Set} the iteration counter $\ell = 1$ \\ 
  \WHILE{$\ell \leq J$ and
  $\sum_{i\in\mathcal{I}}\theta_i > 0$}
  \STATE \textit{Set} $\mathcal{T} = \{24(\ell-1)+1, ..., 24\ell\}$
  \STATE \textit{Set} $\mathcal{D} = \{\ell+1, ..., J\}$
  \STATE \textit{Input} $\{\mbox{V}_{t,i}, \mbox{H}_{t,i}\}_{t \in \mathcal{T}, i \in \mathcal{I}}$ and $\{\mbox{V}^L_{d,i}, \mbox{H}^L_{d,i}\}_{t \in \mathcal{D}, i \in \mathcal{I}}$
  \STATE \textit{Evaluate} $\{f_{t,i}\}_{t \in \mathcal{T}, i \in \mathcal{I}}$ and $\{\alpha_{t,i}\}_{t \in \mathcal{T}, i \in \mathcal{I}}$
    \STATE \textit{Evaluate} $\{f^L_{d,i}\}_{d \in \mathcal{D}, i \in \mathcal{I}}$ and $\{\alpha^L_{d,i}\}_{d \in \mathcal{D}, i \in \mathcal{I}}$
  \STATE \textit{Solve} the MILP model
  \STATE \textit{Return} the optimal solution of the $\ell$-th run, $\mathcal{S^*}^\ell = \{\mathcal{S}^{*^\ell}_{\text{STH}}, \mathcal{S}^{*^\ell}_{\text{LTH}}\}$
  \FOR{$i=1 \in \mathcal{I}$}
  \IF{$\sum_t(m_{t,i}^\ell+n_{t,i}^\ell)>0$}
  \STATE $\theta_i = 0$ \textit{(turbine no longer requires maintenance)}
  \ENDIF
  \ENDFOR
  \STATE $\ell= \ell+1$
  \ENDWHILE
 \RETURN \hspace{-0.25cm} the final solution $\mathcal{S}^* = \{\mathcal{S}^{*^1}_{\text{STH}},...,\mathcal{S}^{*^{\ell-1}}_{\text{STH}}\}$
 \end{algorithmic}
 \end{algorithm}

\section{Numerical Experiments}
This section discusses the data used in our case studies and the setup of our numerical experiments. 

\subsection{Data description and processing}
Wind speed and wave height data are obtained from the E06 Hudson South floating LiDAR buoy recently deployed by NYSERDA in the New York/New Jersey Bight \cite{nyserda2019}. The buoy has been deployed in proximity to at least three ongoing offshore projects which, cumulatively, will add about $2.8$ GW to the U.S. offshore wind capacity by $2024$. The data used in this study spans a total of $2,400$ hours from September $2019$ to January $2020$ and include the time series of the hub-height wind speeds. We utilize this data to simulate turbine-specific wind speed time series at multiple spatial locations by drawing random samples at each time point from a normal distribution with the buoy's recorded wind speed as its mean and a standard deviation of $1$ m/s, chosen to mimic the within-farm spatial variability as reported in prior studies \cite{ezzat2019spatio}. 

For speed-to-power conversion, a publicly available dataset from a U.S. wind farm is used, wherein the normalized power level of several turbines is provided as a fraction of each turbine's rated capacity \cite{ding2019data}. We use this data to estimate each turbine's power curve using the method of bins, which is the recommended approach by the International Electrotechnical Committee (IEC) standard \cite{international2013iec}. The estimated turbine-specific power curves are then used to determine the hourly and daily normalized power values, $f_{t,i}$ and $f_{d,i}^L$, respectively, given turbine-specific wind speed time series $V_{t,i}$ and $V_{d,i}^L$. Those, combined with $R$, will be used to estimate the power level of a turbine (in MW) using (\ref{eq:eqnom}) and (\ref{eq:nom2}). 

Hourly electricity price data are obtained from the open data repository of PJM \cite{PJM}, a regional transmission organization (RTO) that covers 13 states in the eastern U.S. The selected pricing node, Jersey Central Power \& Light (JCP\&L), services parts of the east coast of New Jersey, where future offshore wind farms are set to be installed.

\subsection{Experimental setup} 
We consider an offshore wind farm with $I = 10$ turbines, each with rated capacity of $R = 12$ MW (following General Electric (GE)'s $12$-MW Haliade-X offshore wind turbine \cite{GE}). Two case studies are conducted. In the first case study, we simplify the model by assuming no curtailment (i.e., $C_t = 1, ~\forall t \in \mathcal{T}$), the same repair time for all wind turbines (i.e., $\tau_i = \tilde{\tau}, ~\forall i \in \mathcal{I}$) and a constant electricity selling price (i.e., $\Pi_d = \Pi_t = \tilde{\Pi}, ~\forall t \in \mathcal{T}, d \in \mathcal{D}$). The motivation behind those simplifications is to ensure that the difference in the final schedules and performance of various scheduling strategies is solely attributed to the impact of accessibility, production loss, and crew dispatch. In the second case study, we relax those assumptions; we use variable electricity prices and repair times, and a constant curtailment of $2\%$, i.e., $C_t = 0.98$, following the annual PJM trend for wind curtailment \cite{bird2016wind}. In what follows, we discuss key parameter selections. 

\subsubsection*{Electricity price} For the electricity price of the first case study, we use the non-incentivized levelized cost of energy (LCOE) for offshore wind projects built after $2018$ in the U.S., which is about $\$80$ MWh$^{-1}$ \cite{wiser20192018}. 
For the second, we use actual electricity price data from PJM described in Section III-A with a federal production tax credit (PTC) of $\$24$ MWh$^{-1}$ \cite{wiser20192018}.

\subsubsection*{Maintenance costs and parameters}
Following \cite{yildirim2017integrated}, we set the cost of PM and CM tasks as K $ = \$4,000$ and $\Phi = \$16,000$, respectively. Maintenance crews are paid by the hour; we assume a flat cost rate of $\Psi = \$250\hspace{0.5mm}\mbox{h}^{-1}$ per crew, with an overtime cost rate of $\mbox{Q} = \$125\hspace{0.5mm}\mbox{h}^{-1}$ \cite{maples2013installation}. We assume that there are $\mbox{B} = 2$ maintenance crews available, each consisting of $2$ technicians. We set $\mbox{W} = 8$ hours to be the maximal number of hours with standard payment, after which overtime costs will be incurred. For the first case study, repair times are assumed to be the same across all turbines, i.e., $\tau_i = \tilde{\tau} = 8$ hours. In the second case study, we generate repair times assuming a normal distribution with mean $7.5$ hours (as inspired by \cite{dinwoodie2015reference} for minor repairs), and standard deviation of $3$ hours. This resulted in a minimal sampled repair time of $\sim$\hspace{-0.25mm}$2$ hours, which typically corresponds to a simple manual reset repair. Vessels are rented at a daily rate of $\Omega = \$2,500$/day, corresponding to medium-sized CTVs \cite{dalgic2015offshorelogistics, thomsen2014offshore}. 

\subsubsection*{Accessibility}
CTVs have maximum wave height safety limits, $\eta$ (m), above which, it is unsafe for the vessels to dock on the platform in the base of the offshore turbine. In the U.S., these limits are borrowed from those in the offshore oil \& gas industry, ranging from $0.50$ m to $1.30$ m \cite{maples2013installation}, depending on the vessel's size. In the EU, the corresponding wave height safety limits range from $1.20$ m to $1.75$ m \cite{thomsen2014offshore}. Access to the turbine's nacelle is also constrained by an upper safety limit of wind speed, $\nu$ (m/s), which approximately is around $15$ m/s \cite{windspeedlim2017}. For this study, we set $\eta = 1.5$ m and $\nu =  15$ m/s.

\subsubsection*{Residual life estimations}
RLEs are provided in days, and in both case studies, we assume RLEs to be uniformly placed over the optimization horizon in $5$-day increments (i.e., $5, 10,\hdots ,50$). For the first case study, two unexpected failures for WTs $1$ and $3$, are assumed to take place at the $17$th and $36$th day, respectively. No unexpected failures are considered in the second case study. 

\subsubsection*{Evaluation}
The optimization horizon is set at $J = 60$ days, wherein the rolling-horizon procedure described in Section II-C will be used to obtain a final hourly schedule $\mathcal{S}^*$, which will represent one data point (or scenario) for comparison. We then shift the whole optimization horizon by $1$ day to obtain a new scenario, and repeat the whole procedure described in Algorithm 1. 
Given our $2400$-hour data coverage, this corresponds to a total of $30$ weather scenarios (and hence, $30$ solutions). The performance of a maintenance strategy is evaluated, across all $30$ scenarios, in terms of several O\&M metrics (to be defined in Section IV). In addition to HOST, the following prevalent scheduling strategies are considered: 
        
        \emph{(1) Corrective Maintenance Strategy (CMS)}: Maintenance actions are only scheduled reactively (i.e. post-failure).
        
        \emph{(2) Time-Based Strategy (TBS)}: Maintenance actions are scheduled, whenever feasible, right before the turbines fail. 
    
    \emph{(3) Production-Based Opportunistic Strategy (PBOS)}: A special case of HOST wherein production-based opportunities are prioritized, but not dispatch- or access-based opportunities. 
    
    \emph{(4) Besnard et al. (2009), Opportunistic Strategy (BESN)}: This is akin to the opportunistic framework proposed in  \cite{besnard2009optimization} which accounts for dispatch- and production-based opportunities, but not access-based opportunities. We adapted our own version of this strategy using our formulation in Section II-B.

\section{Results}
We start this section with an in-depth analysis for one weather scenario from case study I to best illustrate the key differences in the schedules obtained from different strategies. The insights derived using this scenario will serve as the basis to justify the results of Section IV-B where the performance across all scenarios are presented for both case studies.

As shown in Section II-B, the proposed model formulation is a mixed integer linear program (MILP), and there are multiple off-the-shelf solvers that use mathematical programming algorithms to find global optimal solutions for MILPs (or solutions that fall within a predefined optimality gap from the global optimum). In this study, the solutions are obtained using the Gurobi version 9.1 solver in Python (which adopts a branch-and-bound algorithm), run on a standard laptop, with a default optimality gap of $0.01$\% \cite{gurobi}. In Table \ref{tab:sol_times}, we report the average solution time for a full hourly schedule (i.e., until all WTs have been visited), and that of a single iteration, which aligns with the practical case where the farm operator's focus is on the hourly day-ahead operations. To demonstrate the scalability of the model, we also report the solution times for cases when the number of WTs is gradually increased. 

\begin{table}[h]
    \caption{Average solution times vs. number of wind turbines}
    \label{tab:sol_times}
    \renewcommand{\arraystretch}{1.2}
    \centering
    \begin{tabular}{c | c | c | c}
    \hline\hline
    \textbf{Number of WTs} & $10$ & $20$ & $30$\\
             \hline
    \textbf{Full Schedule} & $35.2$ sec & $130.9$ sec & $205.2$ sec \\
    \textbf{Single Iteration} & $1.2$ sec & $2.5$ sec  & $4.6$ sec \\
    \thickhline
    \end{tabular}

\end{table}

\subsection{Results from one representative scenario}
\begin{figure}[]
    \centering
\includegraphics[width=7.6cm]{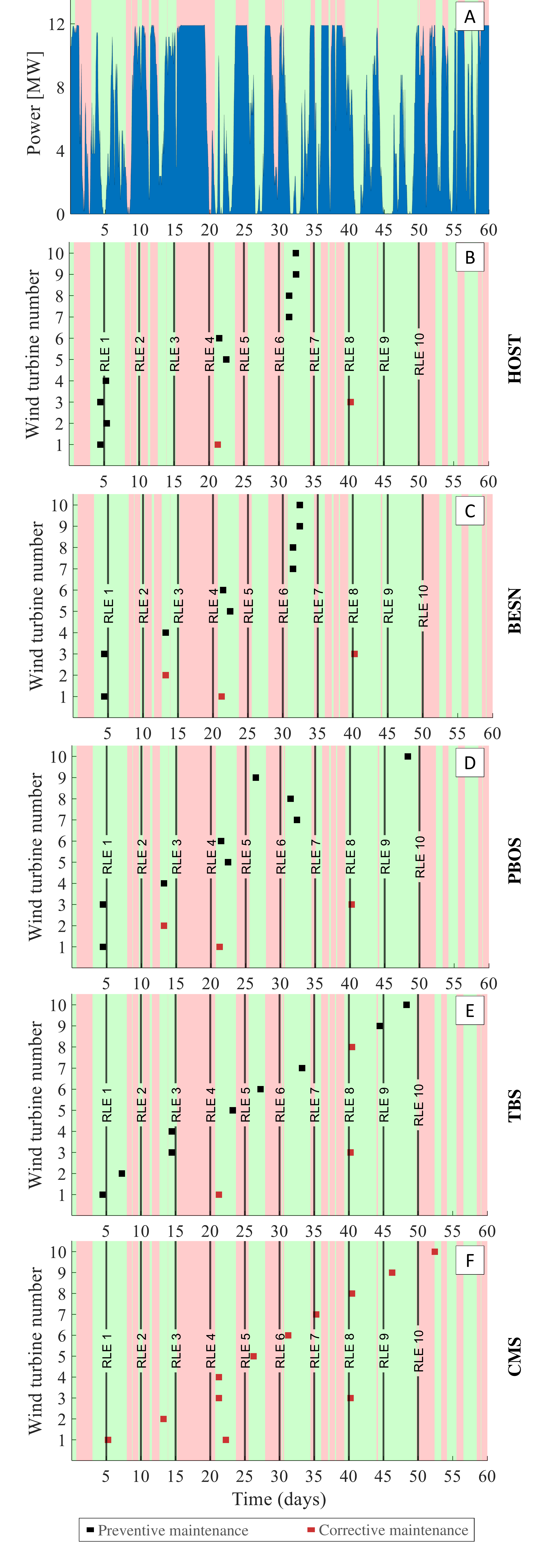}
    \vspace{-0.5cm}
    \caption{Optimization results for one representative scenario. (A): power production levels, averaged across the $10$ WTs. (B)-(F): maintenance schedules obtained using HOST, BESN, PBOS, TBS and CMS. Black and red squares denote the start times of scheduled PM and CM actions, respectively. The background color denotes accessible (green) or inaccessible (red) time periods. }
    \label{fig:cs_weather}
\end{figure}
\begin{figure}[]
    \centering
    \includegraphics[width=8cm, trim = 0 0.25cm 0 0]{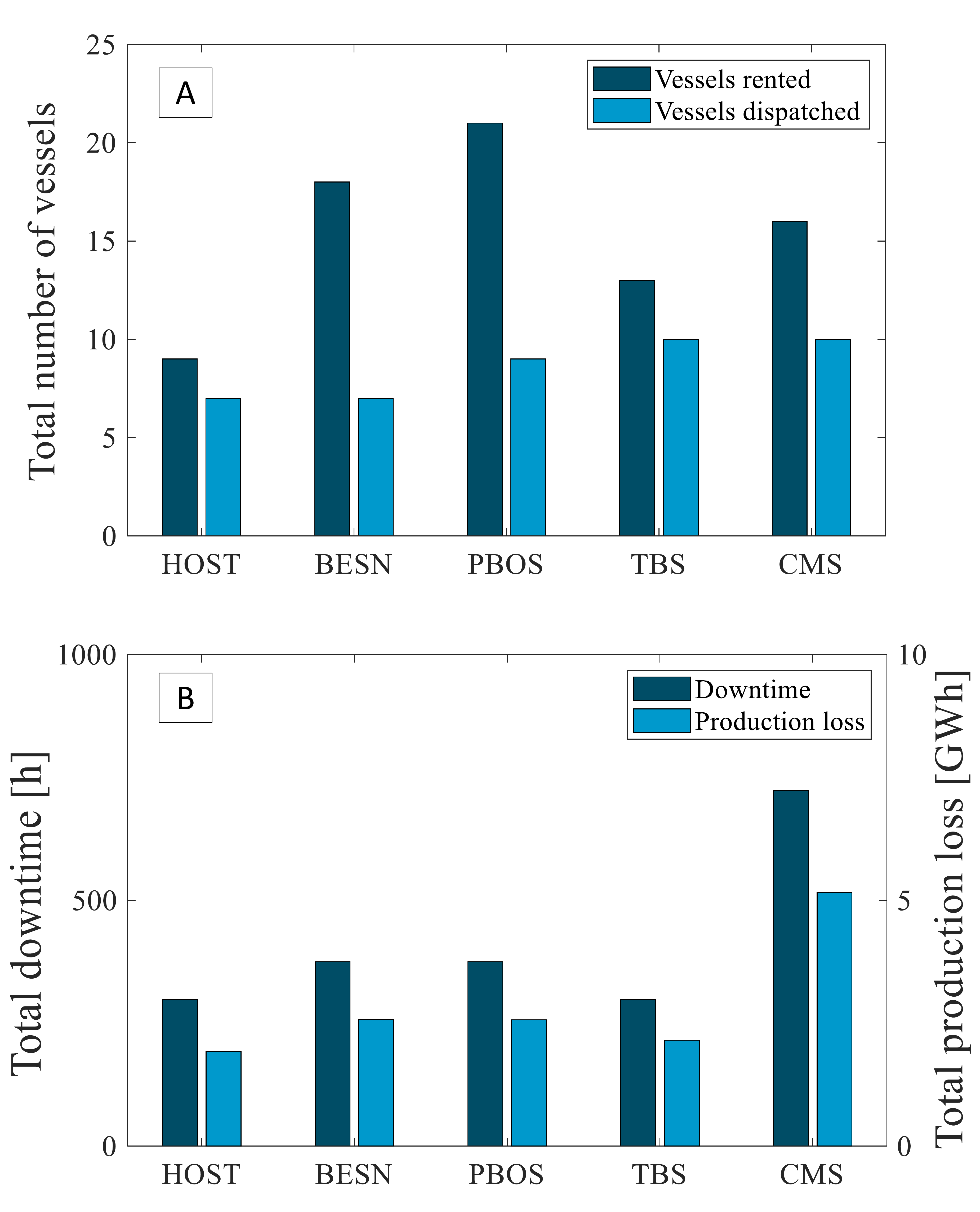}
    \vspace{-0.25cm}
    \caption{(A) Number of vessels rented versus those that were eventually dispatched for each strategy. (B) Total downtime and production losses. Both Figures correspond to the representative scenario of Section IV-A.     }
    \label{fig:cs_powerloss}
\end{figure}

Figure \ref{fig:cs_weather}A shows the average power level across the $10$ WTs for one weather scenario in case study I (Later, we show the full case study results). Figure \ref{fig:cs_weather}B-F illustrate the schedules obtained from all strategies, wherein the $y$-axis represents the WT index, $i = 1, ..., 10$, and the $x$-axis denotes time in days. Black and red squares represent the start times of scheduled PM and CM actions, respectively, while vertical lines show the turbine-specific RLEs. A visual examination of the results presented in Figure \ref{fig:cs_weather} can help us draw the following insights:
    
    \textit{Crew-Dispatch-based opportunities:} First, it is clear that HOST (Figure \ref{fig:cs_weather}B) and BESN (Figure \ref{fig:cs_weather}C), compared to other strategies, group maintenance tasks more aggressively to avoid unnecessary crew dispatch costs. Specifically, both HOST and BESN schedule all maintenance tasks in $7$ days, while the other strategies have more spread-out schedules ($9$+ days). For example, once a PM action has been scheduled for WT1 at day 4, both strategies leverage the opportunity to group that task with other PM tasks. This is further confirmed in Figure \ref{fig:cs_powerloss}A which depicts the number of vessels dispatched for each strategy (light blue bars) showing that HOST and BESN yield the least number of vessel dispatches, compared to strategies that do not incentivize crew-dispatch-based grouping.
    
    \textit{Accessibility-based opportunities:} Figure \ref{fig:cs_weather} shows that access windows can be intermittent, and frequently disrupted by potentially unsafe wind/marine conditions. By ignoring accessibility in the scheduling phase, operators are not incentivized to seize access opportunities when they arise, and hence, additional costs are incurred due to frequent mission aborts/delays wherein an operator schedules a maintenance, but end up delaying it due to access limitations. This drives both an increase in the number of failures and CM tasks (at a higher cost, and only when the weather allows it), as well as in the number of unnecessary vessel rentals which are not dispatched due to access limitations. This is the case for WT2 in the schedules of BESN (Figure \ref{fig:cs_weather}C) and PBOS (Figure \ref{fig:cs_weather}D); a PM action was initially planned in day $8$, but was delayed due to unfavorable sea conditions all the way to day $13$ where a CM task was performed, instead, thus incurring additional CM costs, a total of $3$ days of downtime, notwithstanding the actual time needed for maintenance. This is evident Figure \ref{fig:cs_powerloss}A, showing (with the exception of HOST) sizable unnecessary vessel rentals (dark blue bars) relative to those dispatched.
    
      \textit{Production-based opportunities:} It also clear from Figure \ref{fig:cs_weather}B that HOST schedules all maintenance tasks during periods of low power production to offset downtime-imposed revenue losses. Other strategies result in higher production losses, either because they do not prioritize the minimization of production loss (TBS and CMS) or due to larger turbine downtimes (BESN and PBOS). 
      This is further confirmed in Figure \ref{fig:cs_powerloss}B wherein HOST is clearly shown to yield the minimum turbine downtime and production loss. 

For this representative scenario, HOST results in a total cost of \$$272.6$K, substantially lower than that of BESN (\$$358.6$K), PBOS (\$$366.1$K), TBS (\$$312.8$K), and CMS (\$$653.4$K). Next, we present the exhaustive results for both case studies I and II, across all $30$ scenarios. 

\subsection{Exhaustive results}
\subsubsection{\underline{Results from case study I}}
\begin{figure}[b]
    \centering
    \includegraphics[width=7.5cm, trim = 0 0 0 2.25cm]{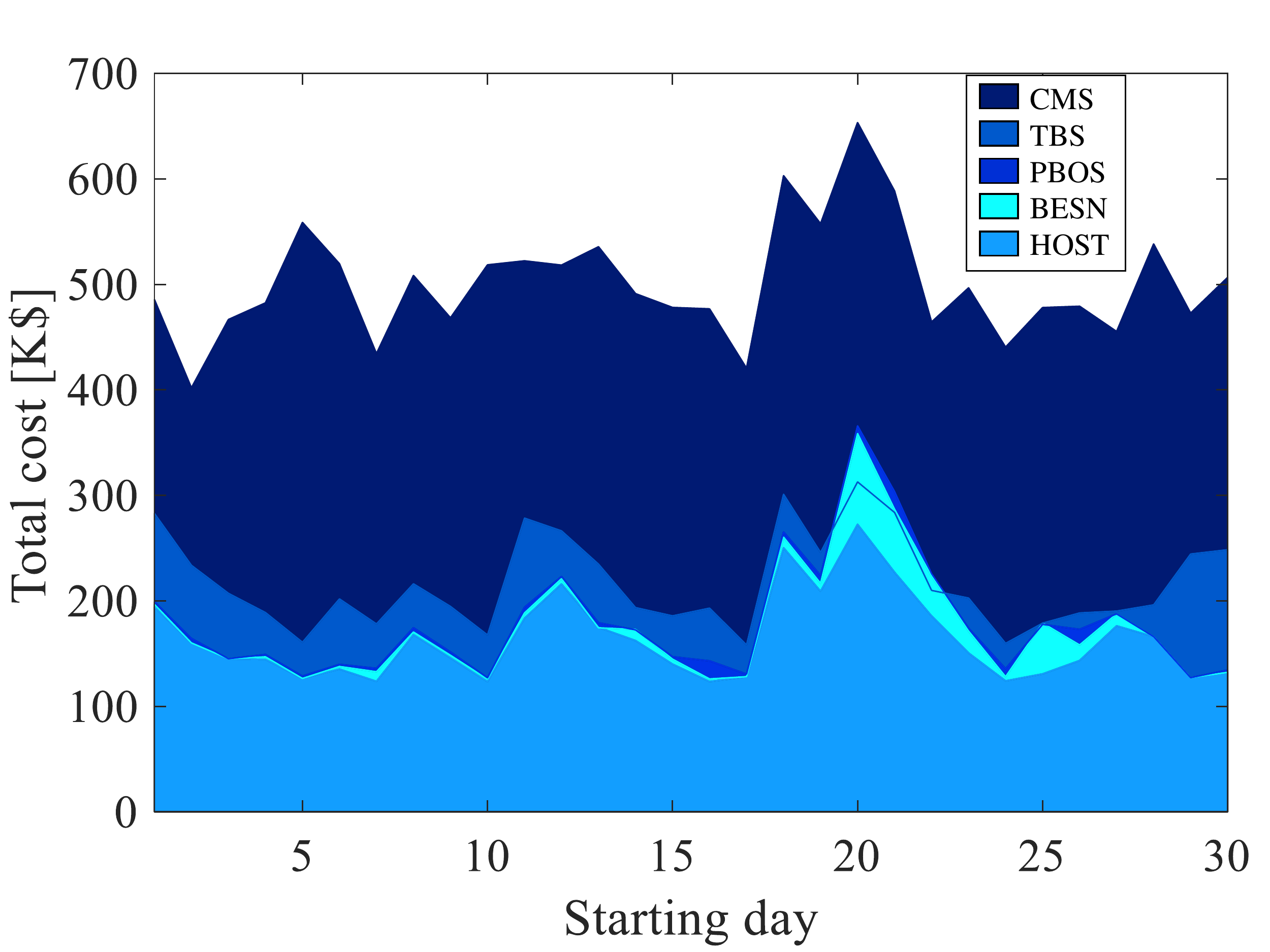}
    \caption{The total costs of the competing strategies, across the $30$ scenarios.}     \label{fig:scenarios_costs}
\end{figure}

\begin{figure}[]
    \centering
   \includegraphics[width=8cm]{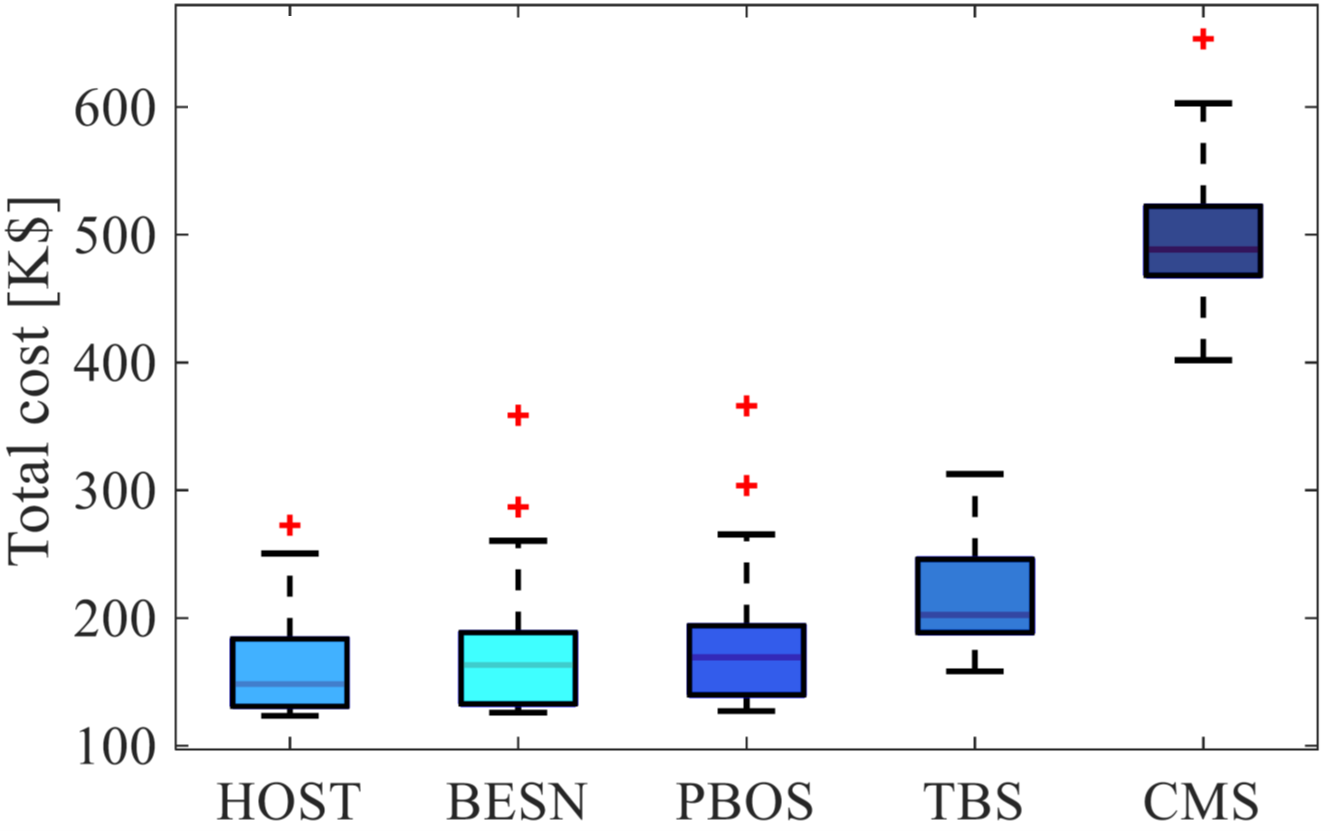}
    \caption{Boxplots of the total costs across the $30$ scenarios for case study I.} 
    \label{fig:bench_boxplots}
    \vspace{-0.25cm}
\end{figure}

Figure \ref{fig:scenarios_costs} shows the total costs of each strategy across the $30$ scenarios for case study I as arranged by starting day, wherein HOST is shown to consistently outperform all other strategies under varying weather scenarios. The corresponding boxplots are shown in Figure \ref{fig:bench_boxplots}. On average, the total costs of HOST for case study I are $6.8$\% lower than BESN, $8.9$\% lower than PBOS, $24.9$\% lower than TBS and $67.4$\% lower than CMS.

The overall performance of HOST is then compared with that of its competitors in terms of key O\&M criteria: the total number of vessel rentals, vessel utilization as defined by the fraction of rented vessels which were eventually dispatched (as a proxy for resource utilization),
total downtime, downtime due to inaccessibility, total farm-level production loss, the number of PM and CM tasks, and lastly, the total costs. The results, presented in Table \ref{tab:metrics} (shaded rows), show substantial improvements for HOST across all relevant O\&M metrics.   

\subsubsection{\underline{Results from case study II}}
Similarly, Table \ref{tab:metrics} (non-shaded rows) summarizes the results of case study II, where it is shown that HOST outperforms the four benchmarks across all metrics, except for the total production losses and vessel utilization, where its performance is comparable to that of PBOS and CMS, respectively. We attribute this slight deterioration in performance to a compromise that HOST makes to reduce vessel rentals and, hence, total costs, as shown in the percentage improvements reported in the last column of Table \ref{tab:metrics}. This reveals HOST's ability to prioritize which ``opportunistic'' aspect is more economically valuable, depending on the available weather and O\&M information. \par 

Note that the removal of the two unexpected failures (which also reduces the total downtime and production losses), power curtailment, and variable electricity prices in case study II, resulted in a considerable difference in the values of some metrics, including total costs, compared to those reported in case study I, yet, the relative performance of the methods and conclusions still hold. On average, the total costs of HOST for case study II are $4.1$\% lower than BESN, $5.2$\% lower than PBOS, $49.5$\% lower than TBS and $77.0$\% lower than CMS.

\begin{table*}[]
    \caption{Results for case studies I and II. Bold-faced values indicate best performance.}
    \label{tab:metrics}
    \renewcommand{\arraystretch}{1.4}
    \centering
    \begin{tabular}{c | c | c c c c c c c c c}
    \hline\hline
    \multicolumn{2}{c}{ } & \thead{\# Vessel \\ rentals} 
    & \thead{Vessel \\ utilization}
    & \thead{Total \\ downtime (h)} & \thead{Accessibility \\ downtime (h)}  & \thead{Production \\ losses (MWh)} & \thead{\# PM \\ ~~actions~~} & \thead{\# CM \\ ~~actions~~} & \thead{Total cost \\ (\$K)} & \thead{\% IMP in \\ total cost (\%)} \\
    \hline
        \rowcolor{lightgray}
    \cellcolor{white}\parbox[t]{2mm}{\multirow{2}{*}{\rotatebox[origin=c]{90}
    {\textbf{HOST}}}}  & {Case Study I} & \textbf{$\mathbf{8.2}$} &  
    \textbf{$\mathbf{83.4}$}\% &
    \textbf{$\mathbf{153.9}$} & \textbf{$\mathbf{57.9}$} & \textbf{$\mathbf{555.7}$} & \textbf{$\mathbf{9.8}$} & \textbf{$\mathbf{2.2}$}  & \textbf{$\mathbf{163.0}$} & - \\
    & {Case Study II} & \textbf{$\mathbf{5.4}$} &  
    \textbf{$83.9$}\% &
    \textbf{$\mathbf{109.7}$} & \textbf{$\mathbf{49.7}$} & \textbf{$69.5$} & \textbf{$\mathbf{9.9}$} & \textbf{$\mathbf{0.1}$}  & \textbf{$\mathbf{74.7}$} & -\\
    
    \hline
            \rowcolor{lightgray}
    \cellcolor{white}\parbox[t]{2mm}{\multirow{2}{*}{\rotatebox[origin=c]{90}{\textbf{BESN}}}}  
    & {Case Study I} &  $10.3$  &    
    \textbf{$69.7$}\% &
    $159.6$  &  $63.6$  &    $611.5$  &     $9.7$  &    $2.3$  & $174.8$ &   $6.8$\% \\
     & {Case Study II} &  $5.6$  &    
    \textbf{$78.7$}\% &
    $113.7$  &  $53.7$  &    $86.3$  &     $9.8$  &    $0.2$  & $77.9$  &  $4.1$\%  \\
    
    \hline
            \rowcolor{lightgray}
    \cellcolor{white}\parbox[t]{2mm}{\multirow{2}{*}{\rotatebox[origin=c]{90}{\textbf{PBOS}}}}  
    & {Case Study I} &  $11.8$   &   
    \textbf{$70.8$}\% &
    $159.6$  &  $63.6$ &    $601.9$   &    $9.6$  &   $2.4$  & $178.9$ &  $8.9$\% \\
    & {Case Study II} &  $7.0$   &   
    \textbf{${88.9}$}\% &
    $115.0$  &  $55.0$ &    $\mathbf{60.2}$   &    $9.8$  &   $0.2$  & $78.8$   &   $5.2$\% \\
   
    \hline
            \rowcolor{lightgray}
   \cellcolor{white}\parbox[t]{2mm}{\multirow{2}{*}{\rotatebox[origin=c]{90}{\textbf{TBS}}}}  
    & {Case Study I} &  $14.2$   &   
    \textbf{$80.0$}\% &
    $164.6$  &  $68.6$ &   $850.7$  &  $8.6$     & $3.4$   & $216.9$ &   $24.9$\% \\
    & {Case Study II} &  $12.4$   &   
    \textbf{$47.3$}\% &
    $147.9$  &  $87.9$ &   $157.3$  &  $5.7$     & $4.3$   & $147.9$ &   $49.5$\% \\
    
    \hline
            \rowcolor{lightgray}
   \cellcolor{white}\parbox[t]{2mm}{\multirow{2}{*}{\rotatebox[origin=c]{90}{\textbf{CMS}}}}  
    & {Case Study I} &  $14.6$  &    
    \textbf{$72.5$}\% &
    $484.0$   &  $388.0$  &    $3,228.9$   &   $0.0$     &   $12.0$  & $500.7$  &   $67.4$\%\\
    & {Case Study II} &  $11.1$  &    
    \textbf{$\mathbf{89.5}$}\% &
    $553.5$   &  $493.5$  &    $2,019.7$   &   $0.0$     &   $10.0$  & $324.2$ & $77.0$\% \\
    \thickhline
    \end{tabular}
\vspace{-0.25cm}
\end{table*}

\subsubsection{\underline{Final discussions}}
We postulate that the substantial improvements in case studies I and II are attributed to:

    \emph{(i) Crew-dispatch-based opportunities:} HOST aggressively groups maintenance tasks whenever economically justifiable and feasible, resulting in significant improvements in ``offshore trips,'' as measured by vessel rentals and utilization. This aspect is extremely relevant in offshore maintenance where dispatch/transport operations have been shown to account for more than $70$\% of total O\&M costs \cite{dalgic2015offshorelogistics, carroll2016failure}.  
    
    \emph{(iii) Access-based opportunities:} A failing offshore turbine may not be accessible for sustained periods of time due to inclement weather and marine conditions ($53\%$ of its operational time, and up to $6$ consecutive days, as shown by our analysis). Missed maintenance opportunities, combined with frequent mission aborts, can inflict substantial O\&M expenditures. Our analysis shows that HOST can offer substantial reductions in downtime by carefully leveraging the available access information at the time of maintenance scheduling. This is particularly evident by the cost reductions achieved by HOST relative to its closest competitor, BESN, which overlooks access-based opportunities. As offshore farms continue to be placed in more remote locations and greater altitudes, formally accounting for this aspect is projected to play an increasingly critical role in offshore maintenance \cite{gilbert2020probabilistic}.
        
    \emph{(ii) Production-based opportunities:} HOST groups maintenance tasks during periods of projected low production and maximal hourly market prices, resulting in the least production revenue losses. To further demonstrate the relevance of considering those hourly-scale production opportunities, we perform a sensitivity analysis for a $3$ WT-case, each requiring $7$-hour maintenance, wherein all other maintenance and market parameters are assumed the same as in case study II. We then compare the optimal schedule against all sub-optimal schedules obtained by shifting the start time of the maintenance tasks to different hourly slots within a day. Two sets of comparisons are performed: (i) when access constraints are relaxed (\# of feasible solutions per day $\approxeq 5000$), and (ii) when access constraints hold (\# of feasible solutions vary per day depending on access states).
    
\begin{figure}[]
    \centering
    \includegraphics[width=\linewidth, trim= 0 1cm 0 0]{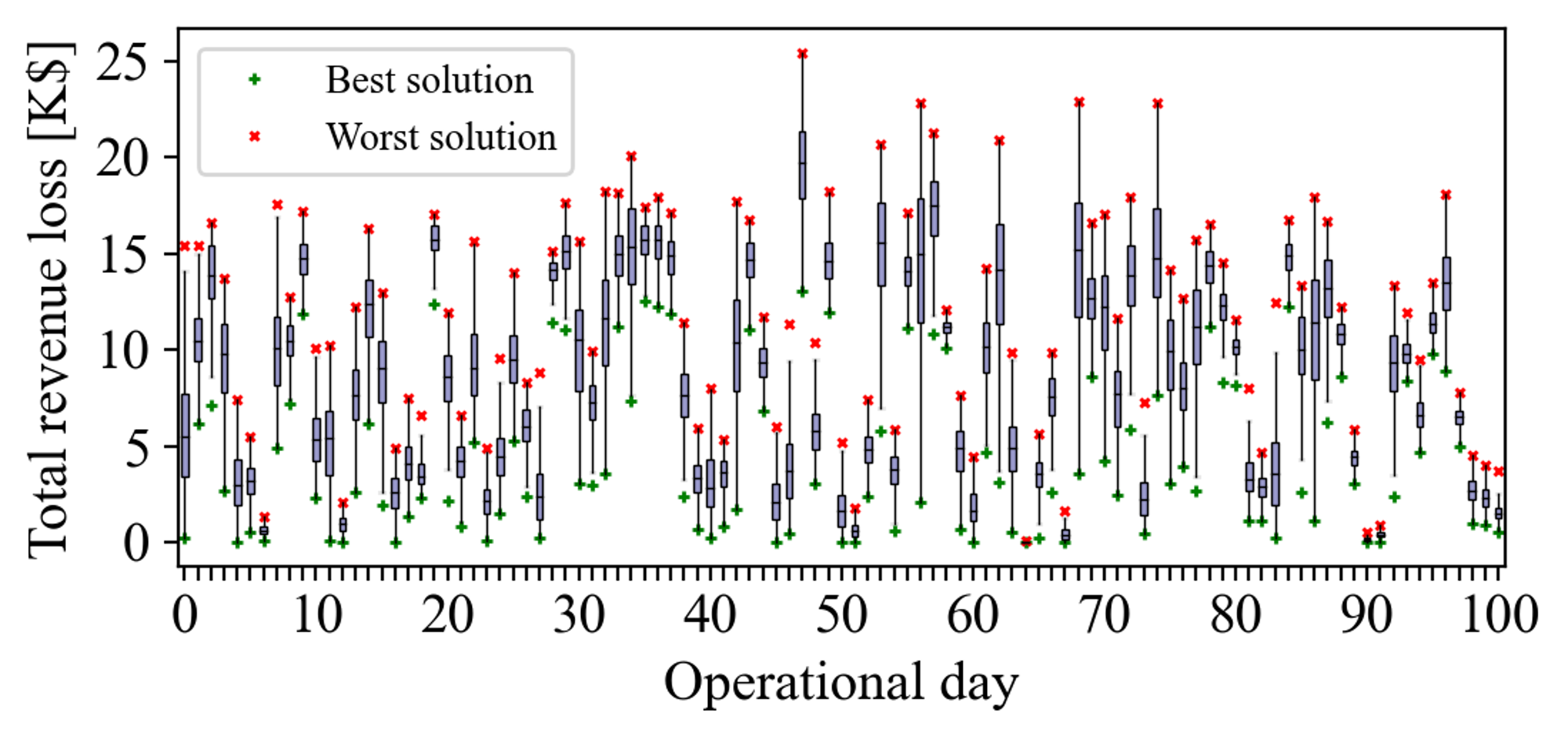}
    \caption{Daily revenue losses due to deviations in maintenance start time relative to the optimal schedule (access constraints relaxed). The large variations demonstrate the relevance of hourly production-based opportunities.} 
    \label{fig:hourly_scale_sens_an}
\end{figure}

    Figure \ref{fig:hourly_scale_sens_an} shows the variation in daily revenue losses due to deviation off the optimal schedule for the first comparison set (access constraints relaxed), while Table \ref{tab:sens_analysis} shows the key revenue loss statistics incurred for both comparison sets. The results in Figure \ref{fig:hourly_scale_sens_an} and Table \ref{tab:sens_analysis} confirm the significance of considering hourly production-based opportunities\textemdash with just $3$ WTs, the revenue losses can increase by as much as $\sim$\hspace{-0.1mm}$\$21$K due to hourly deviations off the optimal maintenance schedule. This production-based dimension of opportunity is especially relevant with the emergence of ultra-scale turbines ($\geq$ 12 MW) \cite{GE,golparvar2021surrogate}, where production loss mitigation is expected to be pivotal to ensuring profitable offshore wind operations.

\begin{table}[]
    \caption{Key statistics of revenue losses due to deviations off the optimal schedule. MAD $\coloneqq$ Max. absolute deviation, $Q_1$, $Q_2$, $Q_3 \coloneqq$ Quantiles of absolute deviation, NMAD $\coloneqq$ Normalized mean absolute deviation (normalized by the range).}
    \label{tab:sens_analysis}
    \renewcommand{\arraystretch}{1.4}
    \centering
    \begin{tabular}{l  c  c }
    \hline
    \hline
        & \textbf{Access constraints relaxed} & \textbf{Access constraints imposed} \\
        \hline
    {MAD} & $\$20.8$K & $\$18.5$K \\
    $Q_1$ & $\$4.8$K & $\$1.4$K \\
    $Q_2$ & $\$6.9$K & $\$4.4$K \\
    $Q_3$ & $\$10.4$K & $\$7.1$K \\
    NMAD & $51.0\%$ & $42.9\%$ \\
    \thickhline
    \end{tabular}
\vspace{-0.25cm}
\end{table}

All of the above results attest our conjecture\textemdash Without a tailored strategy that can adapt itself to the offshore-specific operational and environmental considerations, substantial O\&M cost reductions in offshore wind farms may be forfeited. 

\section{Conclusions \& Future Directions}
With O\&M costs largely defining offshore wind's economic outlook, this paper proposes a holistic opportunistic strategy (HOST) for scheduling offshore maintenance, based on a multi-staged rolling-horizon mixed integer linear formulation, which faithfully aligns with the industrial practice in the offshore wind industry. Exhaustively tested on real-world wind, wave, and power data, HOST is found to consistently yield the cost-minimal solution relative to a set of prevalent maintenance strategies, mainly owing to its ability to fully seize access-, production-, and crew-dispatch-based opportunities.

We envision this work to serve as an anchor point for future research by relaxing some of its potential limitations, including the currently deterministic nature of HOST. In reality, several parameters used herein are inherently stochastic (e.g., wind, wave, access forecasts, RLE predictions, etc.). Future research can therefore investigate formulations and solution methods that formally account for environmental and operational uncertainty, which can further open doors to additional ``dimensions'' of opportunities. For instance, opportunity can be defined based on predicted degradation levels, instead of considering deterministic RLEs. Expanding the current formulation to account for diverse types of turbine components and maintenance actions can further enhance the usability of the proposed framework, since different modes of failure and component interdependencies can be formally accounted for. Finally, consideration of within-farm vessel routing and logistics could provide valuable insights to the problem of offshore maintenance scheduling, especially as offshore wind farms continue to expand in size and capacity.

\appendix  
The nomenclature for the variables, parameters, and sets used in this paper is presented in Table \ref{tab:nom}. The data, codes, and computational environment to reproduce the results of this research are available at \url{https://github.com/petros-pap/HOST}.


%



\section*{Acknowledgment}
This research has been partly supported by the Rutgers Energy Institute (REI), the Rutgers Research Council Grant Program, and the National Science Foundation (NSF) under Grant ECCS-2114422. 


\ifCLASSOPTIONcaptionsoff
  \newpage
\fi



%

\begin{table}[h!] 
    \centering
    \caption{Nomenclature}
    \renewcommand{\arraystretch}{1.2}
    \begin{tabular}{c  c  l}

     \hline\hline
     \textbf{Notation} & \textbf{Domain} & \textbf{Definition}\\
     \hline
     \multicolumn{3}{c}{ \textit{Sets} }\\ 
     \hline
     $i\in \mathcal{I}$ & \{$1, ... ,I$\} & Wind turbines \\
     $t\in \mathcal{T}$ & \{$t_{n-23}, ... ,t_{n}$\} & Hours in short-term horizon\\
     $d\in D$ & \{$d_m, ... ,d_k$\} & Days in long-term horizon \\
     
     \hline
     \multicolumn{3}{c}{ \textit{Decision Variables} }\\ 
     \hline
     $m_{t,i}$ & \{0,1\} & PM is scheduled for WT i at hour $t$\\
     $m_{d,i}^L$ & \{0,1\} & PM is scheduled for WT i at day $d$ \\
     $n_{t,i}$ & \{0,1\} & CM is scheduled for WT i at hour $t$\\
     $n_{d,i}^L$ & \{0,1\} & CM is scheduled for WT i at day $d$ \\
     
     \hline
     \multicolumn{3}{c}{ \textit{Auxiliary Variables} }\\ 
     \hline
     $s$ & $\mathbb{R}$ & Profit for the STH (\$)\\
     $l_{d}$ & $\mathbb{R}$ & Profit for day $d$ of the LTH (\$)\\
     $p_{t,i}$ & $\mathbb{R}^+$ & Hourly power output in the STH (MWh)\\
     $p_{d,i}^L$ & $\mathbb{R}^+$ & Daily power output in the LTH (MWh)\\
     $q$ & $\mathbb{Z}^+$ & Overtime hours worked in the STH (h)\\
     $x_{t,i}$ & \{0,1\} &  WT i is under maintenance during $t$\\
     $y_{t,i}$ & \{0,1\} &  Availability of WT i during $t$ \\
     $y_{d,i}^L$ & \{0,1\} &  Availability of WT i at day $d$ \\
     $v$ & \{0,1\} & Vessel is rented for the STH \\
     $v_d^L$ & \{0,1\} & Vessel is rented for day $d$ of the LTH \\
     
     \hline
     \multicolumn{3}{c}{ \textit{Parameters} }\\
      \hline
      $\Pi_t$  & $\mathbb{R}^+$ &  Hourly electricity price (\$$\cdot\mbox{MWh}^{-1}$)\\
      $\Pi_d$  & $\mathbb{R}^+$ &  Daily electricity price (\$$\cdot\mbox{MWh}^{-1}$)\\
      K  & $\mathbb{R}^+$ &  Cost of a PM action (\$) \\
      $\Phi$ & $\mathbb{R}^+$ &  Cost of a CM action (\$) \\
      $\Psi$  & $\mathbb{R}^+$ &  Maintenance crew cost rate (\$$\cdot \mbox{h}^{-1}$)\\
      $\Omega$  & $\mathbb{R}^+$ &  Vessel rental cost (\$$\cdot \mbox{day}^{-1}$)\\
      Q  & $\mathbb{R}^+$ & Cost of extra work hour (\$$\cdot \mbox{h}^{-1}$)\\
      $\mbox{C}_t$ & $[0,1]$ & Fraction of curtailed power \\
      $\theta_i$ & \{0,1\} & WT $i$ requires maintenance \\
      $\lambda_i$ & $\mathbb{Z}^+$ & Residual life estimate of WT $i$ (days)\\
      $\alpha_{t,i}$ & \{0,1\} & WT $i$ is accessible during $t$ \\
      $\alpha_{d,i}^L$ & \{0,1\} & WT $i$ is accessible at day $d$ \\
      $f_{t,i}$ & [0,1] & Normalized power level during $t$ for $i$ \\
      $f_{d,i}^L$ & [0,1] & Normalized power level at day $d$ for $i$ \\
       B  & $\mathbb{Z}^+$ &  Number of maintenance crews \\
       W  & $\mathbb{Z}^+$ &  Work hours with standard payment (h) \\
       R & $\mathbb{R}^+$ & WT nominal capacity (MW)\\
      $\tau_i$ & $\mathbb{Z}^+$ &  Repair time for WT $i$ (h)\\
       M & $\mathbb{R}^+$ & Arbitrary large number\\
      $\beta$ & $\mathbb{R}^+$ & Arbitrary small number\\
       $\mbox{V}_{t,i}$  & $\mathbb{R}^+$ &  Wind speed in WT $i$ during $t$ (m$\cdot \mbox{s}^{-1}$)\\
       $\mbox{H}_t$  & $\mathbb{R}^+$ &  Significant wave height during $t$ (m)\\
       $\mbox{V}_{d,i}^L$  & $\mathbb{R}^+$ &  Wind speed in WT $i$ at day $d$ (m$\cdot \mbox{s}^{-1}$)\\
       $\mbox{H}_d^L$  & $\mathbb{R}^+$ &  Significant wave height at day $d$ (m)\\
       $\nu$ & $\mathbb{R}^+$ & Wind speed safety limit (m$\cdot \mbox{s}^{-1}$)\\
       $\eta$ & $\mathbb{R}^+$ & Wave height safety limit (m)\\
      
       \thickhline
       \vspace{-.75cm}
       \label{tab:nom}
    \end{tabular}
\end{table}


\printbibliography





\newpage
\begin{IEEEbiography}[{\vspace*{-.75cm}\includegraphics[width=1in,height=1.25in,trim = {0 0cm 0 0cm}, clip, keepaspectratio]{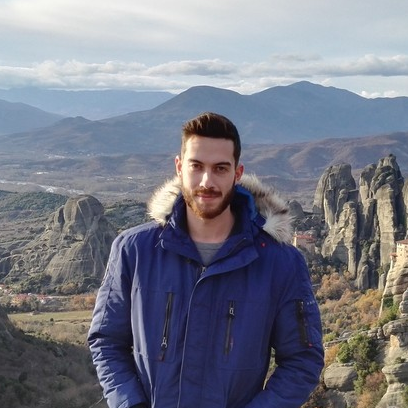}}]{Petros Papadopoulos}
received his BSc and MEng degree in Chemical Engineering from the University of Patras, Greece in 2018. He is currently pursuing a PhD in Industrial \& Systems Engineering at Rutgers University in New Jersey, USA. His research interests are in the areas of operations research and machine learning, applied to renewable energy systems. 
\end{IEEEbiography}

\vskip -3.5\baselineskip plus -1fil

\begin{IEEEbiography}[{\includegraphics[width=1in,height=1.25in, clip,keepaspectratio]{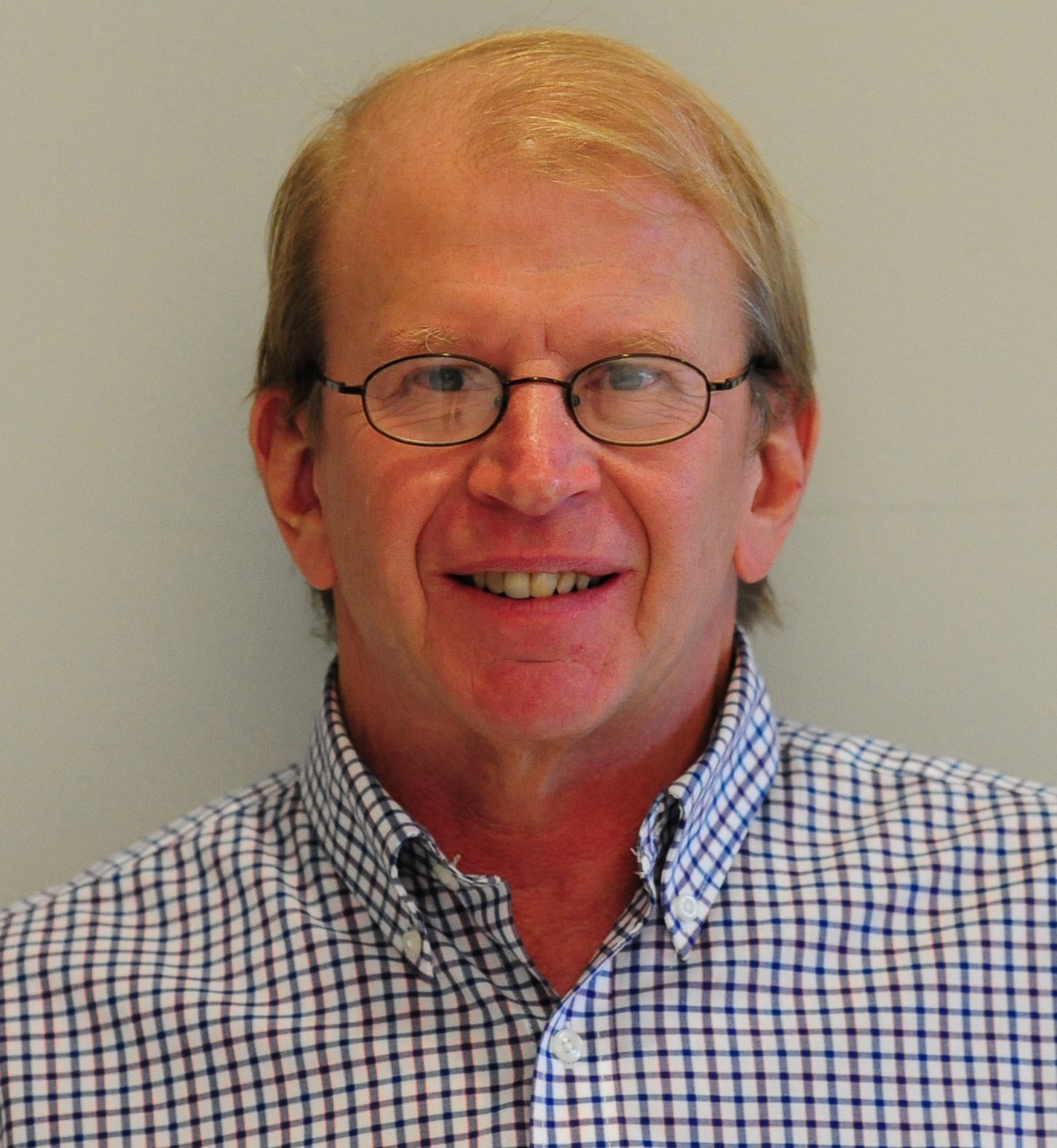}}]{David Coit}
is a professor in the Department of Industrial and Systems Engineering at Rutgers University. His research interests are in system reliability optimization and energy systems optimization. His research has been funded by National Science Foundation (NSF), U.S. Navy, U.S. Army and industry. He had over 120 published journal papers and 90 peer-reviewed conference papers. He has a B.Sc. degree in Mechanical Engineering from Cornell University, MBA from Rensselaer Polytechnic Institute, and M.Sc. and Ph.D. degree from University of Pittsburgh. He is a member of IISE, INFORMS and IEEE.
\end{IEEEbiography}

\vskip -2.5\baselineskip plus -1fil

\begin{IEEEbiography}[{\includegraphics[width=1in,height=1.25in,clip,keepaspectratio]{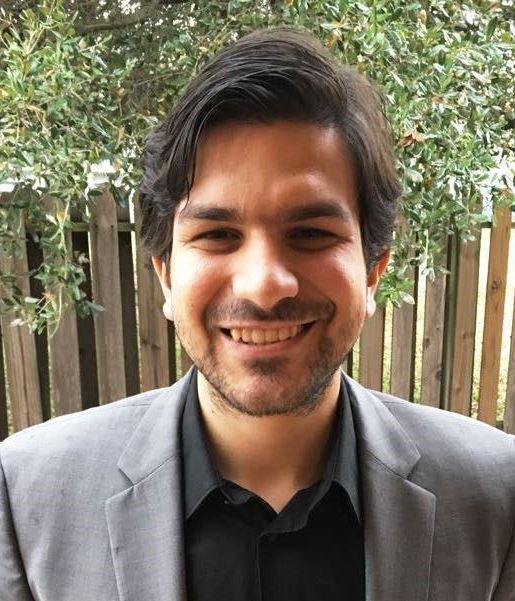}}]{Ahmed Aziz Ezzat}
received his Ph.D. in Industrial \& Systems Engineering from Texas A\&M University in $2019$, and his B.Sc. and M.Sc. degrees in Industrial \& Management Engineering from the Arab Academy for Science and Technology in Alexandria, Egypt, in $2013$, and $2016$, respectively. He is currently an Assistant Professor of Industrial \& Systems Engineering at Rutgers University, where he leads the Renewables and Industrial Analytics (RIA) group which conducts research on spatio-temporal learning and forecasting, quality and reliability engineering, data and decision sciences, with applications to energy analytics and materials informatics. He is a member of IEEE-PES, IISE, and INFORMS.
\end{IEEEbiography}




\end{document}